\documentstyle{article}
\begin{document}
\begin{flushright}
SPIN-1999/28\\
October 1999
\end{flushright}

\begin{center}
{\Large \bf U-duality of Born-Infeld Theory} \\

\vspace{4mm}

{\bf Gysbert Zwart}\\
Spinoza Institute\\
Leuvenlaan 4\\
3584 CE Utrecht\\
and\\
Institute for Theoretical Physics\\
University of Utrecht\\
Princetonplein 5\\
3584 CC Utrecht\\
The Netherlands\\
\end{center}

\begin{abstract}
String theory compactified on a three-torus possesses an $SL(5,{\bf Z})$
U-duality group. We investigate the realisation of this 
symmetry on the Born-Infeld theory on a three-brane, and discuss a 
U-duality covariant formulation of the BPS sector of the theory 
where the rank of the gauge group is treated on an 
equal footing with the fluxes. 
\end{abstract}



\section{Introduction}

The simplest compactifications of string theories are those on tori. 
These preserve maximal supersymmetry, and furthermore they enjoy a large 
discrete symmetry, called U-duality. This U-duality group can be viewed 
as being generated by two distinct sets of symmetries (see e.g. 
\cite{ObersPioline} and references therein). 

The first is T-duality, a perturbative symmetry of string theory, i.e. it 
holds order by order in the string loop expansion. T-duality states that 
strings on circles of radius $R$ are in 
fact equivalent to strings on circles with inverse radii. In the 
identification, the roles of string momentum and winding states are 
interchanged. The T-duality group of string theory on a $d$-dimensional 
torus is $SO(d,d,{\bf Z})$.

The second contribution to the U-duality group follows from the observation 
that type IIA strings on a $d$-torus may alternatively be regarded as 
M-theory on a torus of one dimension higher. This point of view makes 
obvious a geometric symmetry group $SL(d+1,{\bf Z})$ of the torus. From 
the string theory perspective, this is a non-perturbative symmetry: for 
instance, it interchanges non-perturbative D$0$-branes with stringy momentum
modes. 

When these two groups are combined, they generate the U-duality groups, 
whose continuous versions have been known to exist for a long time as the 
hidden 
symmetries of supergravity theories. For various torus dimensions $d$, 
they are listed in table \ref{tbl:uduality}.
\begin{table}[h]
\centerline{
\begin{tabular}{|c|c|}
\hline
$d$ & U-duality group\\
\hline
1 & $SL(2,{\bf Z})$\\ \hline
2 & $SL(2,{\bf Z})\times SL(3,{\bf Z})$\\
\hline
3 & $SL(5,{\bf Z})$ \\ \hline
4 & $SO(5,5,{\bf Z})$ \\ \hline
5 & $E_{6(6)}({\bf Z})$ \\ \hline
6 & $E_{7(7)}({\bf Z})$ \\ \hline
7 & $E_{8(8)}({\bf Z})$ \\ \hline
\end{tabular}
}
\caption{U-duality groups for string theory compactified on various tori}
\label{tbl:uduality}
\end{table}
In the following we will concentrate on the case $d=3$, where the U-duality 
group is $SL(5,{\bf Z})$.

The states in the theory transform in multiplets of the 
U-duality group. In particular this is the case for the $1/2$ BPS states. In 
type IIB strings on a three-torus these states are D$3$-branes, 
three different D$1$-branes (one for each cycle of the torus), three 
fundamental string winding modes and three momenta around the torus.   
These ten objects transform as an antisymmetric tensor under $SL(5)$.
The $1/4$ BPS states, in which we will be mainly interested, are realised as 
combinations of such $1/2$ BPS objects.

The symmetry under U-duality transformations also implies that the 
degeneracies of states related by such a transformation should coincide. 
In the case of $1/4$ BPS states this degeneracy can be calculated in 
perturbative string theory by considering a state with only momentum and 
winding quantum numbers. 
The degeneracy of a state with momentum vector $p_i$ 
and winding numbers $w_i$ is given by $D(p\cdot w)$, with $D(n)$ defined 
by the chiral string partition function:
\begin{equation}
\sum D(n)q^n= 256 \prod \left( \frac{1+q^n}{1-q^n}\right)^8.
\end{equation}
Since any other $1/4$ BPS state can be mapped to such a state, all $1/4$ BPS 
degeneracies should be given by a similar expression.

The purpose of this work (reported in \cite{HVZ,gz}) is to investigate the 
implications of the 
U-duality symmetry of the string theory on the Born-Infeld gauge theory 
living on a 
three-brane wrapping a torus. As we will see, the BPS states of 
the strings have an interpretation in terms of fluxes in the gauge theory 
on the three-brane. We will study the BPS sector of this Born-Infeld 
theory and, via BPS quantisation (following \cite{HV}) determine the 
associated degeneracies of the BPS states in this gauge theory. These 
will turn out to be in accord with the string results. 

\section{The Born-Infeld gauge theory on the three-brane}

We will focus the discussion on the seven-dimensional case, corresponding 
to string theory compactified on a three-torus, with U-duality group 
$SL(5,{\bf Z})$. On the type IIB string theory side, BPS states are built 
from ten distinct objects: three-branes wrapping the three-torus, 
fundamental and D-strings winding around three different one-cycles, and 
momentum modes in the three internal directions. The associated quantum 
numbers transform in the ten-dimensional representation of $SL(5)$.

In the gauge theory on a three-brane wrapping the torus, all these ten 
quantum numbers have an interpretation as fluxes. The relations are given 
in table \ref{tbl:fluxes}. 
\begin{table}[h]
\centerline{
\begin{tabular}{|c|c|}
\hline
string theory & gauge theory\\
\hline
D$3$-brane & rank $N$\\ 
D$1$-branes & magnetic fluxes $B_i$\\
fundamental strings & electric fluxes $E_i$ \\ 
momenta & gauge momenta $P_i$ \\ \hline
\end{tabular}
}
\caption{Translation of string theory quantum numbers into gauge fluxes}
\label{tbl:fluxes}
\end{table}
The number of D$3$-branes is of course related to the rank $N$ of the 
$U(N)$ gauge theory. The magnetic fluxes, the zero modes of the magnetic 
field $B_i=\frac{1}{2}\epsilon_{ijk}F_{ij}$, correspond to D-strings, 
whereas the S-dual electric fluxes take the role of fundamental strings. 
Finally the momenta (i.e.\ the components of the integrated Poynting vector 
$E\wedge 
B$) simply translate to the string theory momenta around the torus. 

To bring out the U-duality properties, it is convenient to organise 
these gauge theory quantities in an antisymmetric five by five tensor 
$M_{ij}$, as follows:
\begin{equation}
\label{eqn:M}
M= \left(\begin{array}{ccccc}
 0 & P_3 & -P_2 & E_1 & B_1\\
-P_3& 0 & P_1& E_2 & B_2\\
P_2 & -P_1 & 0 & E_3 & B_3\\
-E_1 & -E_2 & -E_3& 0 & N\\
-B_1 & -B_2 & -B_3 & -N & 0
\end{array}\right).
\end{equation}
The $SL(5)$ acts on this matrix by conjugation. One can easily recognise 
the two subgroup $SL(3)$ and $SL(2)$. The former is the geometric 
symmetry on the torus and acts as such on the three vectors. It sits in 
the top lefthand block of the $SL(5)$ matrices. The two by two lower 
righthand block realises the $SL(2)$ which is the electromagnetic duality. 
It mixes electric and magnetic components. Both subgroups leave the rank $N$
untouched. We will discuss those transformations affecting the rank in 
the following.

The particular gauge theory on the D$3$-brane that we want to consider  is 
the Born-Infeld gauge theory. The abelian version of its action is given by
\begin{equation}
S_{BI}= \frac{1}{g_s}\int \sqrt{\det (G_{\mu\nu} +F_{\mu\nu})}.
\end{equation}
The inverse string coupling in front of the action is typical of 
D-branes. We omit possible non-trivial $B$-field background contributions.
The generalisation to higher rank is thought to be given by a symmetrised 
trace over the gauge group; this issue is not fully resolved yet however, 
for a discussion see e.g. \cite{Tseytlin}. We will compute the 
Hamiltonian and BPS masses for the abelian case and assume the 
generalised result for arbitrary $N$.

In order to calculate the Hamiltonian we need to introduce the electric 
field 
$$
E^i = \frac{\delta{\cal L}}{\delta A_i}.
$$
It now turns out that the square of the Hamiltonian density $H$ can be 
expressed 
in a simple way in terms of the matrix $M_{ij}$ defined in 
equation (\ref{eqn:M}), as 
\begin{eqnarray}
H^2 &=& \frac{1}{g_s^2}(N^2 + B^2) + E^iG_{ij}E^j + 
P_i(G^{-1})^{ij}P_j\nonumber\\
&=& -\frac{1}{2} \mbox{Tr } M^2.
\end{eqnarray}
In the first line, we see the contribution from the D$3$ and D$1$-branes, 
with the characteristic coupling constant dependence, and then the 
terms corresponding to winding and momentum. The second line demonstrates 
that the Hamiltonian takes a very simple form in terms of the $SL(5)$ 
tensor $M$. (We can absorb the coupling, as well as any non-trivial 
background fields, in a five-dimensional metric). 

This form suggests an $SL(5)$ covariant description  of the theory. 
However, the matrix $M$ does not have arbitrary components, there exist 
relations between them. Remarkably we can write these relations again in 
an $SL(5)$ covariant form, as the constraint
\begin{equation}
K^i = \frac{1}{8}\epsilon^{ijklm}M_{jk}M_{lm} = 0.
\label{eqn:constr}
\end{equation}  
In components the five-vector $K^i$ is given by
$$
K= (NP_i - (E\wedge B)_i, -P\cdot B, P\cdot E).
$$
The first three components are precisely the definition of the Poynting 
vector, while the last two components are automatically zero whenever the 
first three are. 

We are therefore led to consider an arbitrary matrix $M$, provided it 
satisfies the five-vector constraint $K= 1/2(M\wedge M)=0$. At this point a 
major problem is of course that, to make the connection to gauge theory, 
while $E$, $B$ and $P$ may be position dependent, the rank $N$ should of 
course be a constant. We will turn to this in a moment.

Finally, we are interested in the BPS states of the theory. Again the BPS 
masses can be written in a nice form using the matrix $M_{ij}$. The BPS 
mass is a function of the ten charges, which are given by the zero modes 
of $M$. We write these as $m_{ij}= \int M_{ij}$. In terms of this matrix 
of fluxes, the BPS mass formula takes the form
\begin{equation}
M_{BPS}^2= -\frac{1}{2}\mbox{Tr } m^2 + 2 |k|.
\end{equation}
Here $k$ is the zero mode equivalent of $K$, i.e.\ $k=1/2 (m\wedge m)$. Note 
that while the space dependent $K$ is automatically zero, $k$ is not. In 
fact $k=0$ only for $1/2$ BPS states, while $1/4$ BPS states have 
non-zero vector $k$. The BPS equations can be expressed in a covariant 
fashion in terms of $M$ and its zero-modes $m$ as well.

We will go on to quantise the space of BPS states, in order to try to 
determine the quantum degeneracies of the BPS states. We will first 
review the method of BPS quantisation introduced in \cite{HV} for the 
$U(N)$ Yang-Mills theory; then we will apply this to the Born-Infeld case.

\section{BPS quantisation of Yang-Mills theory}

In \cite{HV} Hacquebord and Verlinde discussed the question of $SL(5)$ 
invariance of the BPS spectrum in the context of Yang-Mills theory on a 
torus.
In the Yang-Mills theory we have the fields $A_\mu$, the vector 
potential, and six (adjoint) scalar fields $X^I$. The BPS equations 
depend on the fluxes of the configuration. In the simple case where only 
the momentum in the one-direction, $p_1$, and the rank $n$ are non-zero, 
we may gauge-fix $A_0$ and $A_1$ to zero and then obtain the BPS equations
\begin{equation}
(\partial_0-\partial_1)A_{2,3}=0,\quad (\partial_0-\partial_1) X^I = 0, \quad
[A_i,A_j]=[A_i,X^I]=[X^I,X^J]=0.
\end{equation}
These equations were recognised in \cite{HV} as the left-moving sector of 
a matrix string theory. Due to the vanishing of the commutators, one can 
take all $n$ by $n$ matrices to be diagonal. At first sight this seems to 
imply that we have simply $n$ distinct left-moving theories on a string. 
However, in the periodicity conditions in the coordinate $x_1$ one may 
include a permutation of the eigenvalues,
$$
A_i(x_1 + 2\pi)= SA_i(x_1)S^{-1},
$$
and similarly for the $X^I$, so that effectively one describes the 
``long strings'' introduced in \cite{DVV} as the twisted sectors of a 
conformal field theory on a symmetric product, with a total length $n$.

Hacquebord and Verlinde concentrated on the case where one has just one 
string of length $n$. In this case, quantisation of the theory restricted 
to the BPS configurations yields left-moving oscillators with a 
fractional moding, in multiples of $1/n$. Then, to obtain a total 
momentum $p_1$ one should consider states with oscillator number 
$np_1$, so that  the degeneracy of such states is indeed given by 
$D(np_1)$, the result from string theory. 

For more general quantum numbers the degeneracies were argued to be the 
same in \cite{HV}. The total BPS degeneracies for the Yang-Mills $U(N)$ 
gauge theory therefore respect the U-duality group 
$SL(5,{\bf Z})$, at least in the single long string sector. 

The essential point in this result is that in the subsector of the theory 
respecting the BPS conditions, the configurations reduce to strings. The 
length of these strings equals the rank of the gauge theory. We will now 
try to apply the same arguments to the Born-Infeld gauge theory. In the 
abelian case where the theory is well understood, we will again see the 
reduction to a string theory. For the non-abelian version, we will {\em 
assume} that similarly the equations reduce to those of a matrix string 
theory, so that effectively we can also here use the abelian BPS 
equations. We know that in the limit of large $N$, where the theory is 
adequately described by the Yang-Mills theory, this should be the case, 
but for general $N$ this remains an assumption essential for our result. 

\section{BPS quantisation of the Born-Infeld theory}

For the supersymmetric Born-Infeld case we will now reexamine the 
situation. As 
explained above we will use the BPS equations found from the abelian 
Born-Infeld theory, and assume these to be valid for the non-abelian 
case, with the only alteration that the length of the domain on which the
fields live is multiplied by the rank $n$.

Just as in the Yang-Mills case let us start from the easy case, where 
only the quantum numbers associated to the rank, $n$, and the momentum in 
the one-direction, $p_1$, are non-zero. In this case the BPS equations are
\begin{equation}
E_2=B_3,\quad E_3=-B_2, \quad E_1=B_1=P_2=P_3=0.
\end{equation}
If we insert the expressions for $E_i$ and $B_i$ in terms of the gauge 
field $A_i$ we again find the same equation as in the case of the 
Yang-Mills theory,
$$
(\partial_0-\partial_1)A_i=0,
$$
and we are suppressing the six extra scalars $X^I$.
From the fact that the equations are precisely the same we can of course 
conclude that here we have the same degeneracy as the one found in the 
previous situation. However, in order to be able to generalise to arbitrary 
fluxes, it is convenient to go through the calculation in a little more 
detail.

In order to quantise the theory in a lightcone gauge, we identify the 
electric field field with left-moving string coordinates,
$$
E_i\sim\partial X_i,
$$      
enjoying the appropriate commutation relations
$$
[\partial X(\sigma),\partial X (\sigma')] = i\partial \delta(\sigma-\sigma').
$$
From this relation to a string theory one can compute the degeneracy. 
However, let us step back and try to make the analogy to the string 
theory before the fixing to lightcone gauge. In order to do this we 
propose to identify the lightcone coordinates $\partial X^\pm$ with the
rank, $N$, and the momentum, $P_1$. For the moment we therefore assume 
$N$ to be a real, fluctuating field, whose zero mode is the rank $n$. This 
implies imposing a commutation relation between the two quantities,
\begin{equation}
[N,P_1] = i\partial\delta.
\end{equation}
In this notation, if we write down the constraint $K^i=0$ (equation 
(\ref{eqn:constr})), its only non-trivial component $K^1$ takes the form
\begin{equation}
K^1= \partial X^+\partial X^- - \frac{1}{2}\partial X^i \partial X^i = 0
\end{equation}
which we recognise as precisely the Virasoro constraint! Using the gauge 
symmetry generated by this  
constraint we may now fix $N=\partial X^+$ to be a constant, $n$. This 
then determines $P^1=\partial X^-$ in terms of the other fields as
$$
n P_1 = (E\wedge B)_1.
$$

In conclusion, by introducing an underlying pre-theory, in which the 
rank is allowed to be a fluctuating field, we manage to make contact to a 
string theory before the fixing of lightcone gauge. In this theory the 
constraint $K=0$ is interpreted as the Virasoro constraint. The original 
gauge theory is then identified with the lightcone gauge-fixed version of 
this theory, where the rank is identified with the lightcone momentum.

So far we have only introduced some additional structure, which by fixing 
a gauge we again removed. The usefulness of this additional structure 
becomes 
clear when we consider the generalisation to arbitrary fluxes. To solve 
the problem for general fluxes, let us insert the expression in terms of 
the $\partial X$'s from the previous discussion in the matrix $M_{ij}$:
\begin{equation}
M= \left(\begin{array}{ccccc}
 0 & 0 & 0 & 0 & 0\\
 0& 0 & \partial X^-& \partial X^2 & -\partial X^3\\
 0 & -\partial X^- & 0 & \partial X^3 & \partial X^2\\
 0 & -\partial X^2 & -\partial X^3& 0 & \partial X^+\\
 0 & \partial X^3 & -\partial X^2 & -\partial X^+ & 0
\end{array}\right).
\end{equation}
Now, if $M_{ij}$ satisfies the Born-Infeld BPS equations, then, since 
these equations are covariant, an $SL(5)$ transformed $M'_{ij}$ is a 
solution as well. If we therefore conjugate the matrix $M$ above with an 
appropriate $SL(5,{\bf Z})$ matrix, we obtain a new solution with 
arbitrary new zero-modes of the fields (fluxes). The entries of this new 
$M'$ are all linear combinations of the $\partial X$, so the new electric 
and magnetic fields, as well as momenta and rank, are all functions of 
all the $\partial X$'s. 
In particular, since the new rank $N'$ is not anymore simply $\partial 
X^+$, it is no longer a constant. However, from the previous discussion 
we see that we may remedy this simply by making a gauge transformation 
generated by the constraint $K^i$, to fix the gauge so that again $N'$ is 
a constant. Effectively, by performing a U-duality transformation we have 
gone out of the lightcone gauge, and one has to apply a compensating 
gauge transformation to reach a new configuration with constant rank. 
Since this is only a gauge transformation, it does not of course affect 
the degeneracy, so that this is indeed automatically U-invariant.

\section{Conclusions}

We have seen that Born-Infeld theory in $3+1$ dimensions can be naturally 
written in terms of U-covariant objects: the antisymmetric matrix 
$M_{ij}$, together with the five-vector of constraints $K^i=0$. The 
spectrum of BPS masses, as well as the BPS equations take a covariant 
form in terms of these quantities.

To study the degeneracies of the BPS states, we generalised the BPS 
quantisation applied to Yang-Mills theory in \cite{HV}. In the BPS 
sector, we saw that the theory reduced to a string theory, giving rise to 
the stringy degeneracies $D(n)$. Furthermore, we proposed a theory 
underlying the actual gauge theory in the BPS sector, in which the rank 
is treated on an equal footing as the other fields. In this formulation 
the constraint $K$ was identified with the generator of conformal 
transformations. Fixing this theory to lightcone gauge yields the actual 
gauge theory with constant rank $N$.



\end{document}